\documentclass[12pt,preprint]{aastex}
\usepackage{amssymb}
\newcommand{\ld}{{\ell '}}
\newcommand{\ldd}{{\ell ''}}
\newcommand{\wt}{\widetilde}
\newcommand{\wh}{\widehat}

\begin{document}

\title{ A SEARCH FOR X-RAY REIONIZATION SIGNATURES FROM CROSS-CORRELATION OF \emph{WMAP} AND \textbf{ROSAT}
RASS DATA}

\author{Quan Guo\altaffilmark{1,2}, Xiang-Ping Wu\altaffilmark{1},
HaiGuang Xu\altaffilmark{3} and JunHua Gu\altaffilmark{3}}

\altaffiltext{1}{National Astronomical Observatories,
Chinese Academy of Sciences, Beijing 100012, China}
\altaffiltext{2}{Graduate School of Chinese Academy of Sciences, Beijing 100049,
China}
\altaffiltext{3}{Department of Physics, Shanghai Jiao Tong University,
800 Dongchuan Road, Shanghai 200240, RPC}

\begin{abstract}
We present an observational search for the possible correlation
between cosmic microwave background (CMB) polarization map
and soft X-ray background (SXRB) based on the ROSAT All-sky Survey (RASS)
archive and WMAP
five-year observations. This is motivated
by the fact that some of the CMB polarization may arise from
the scattering of CMB photons due to the free electrons generated by
X-ray heating sources in the epoch of reionization. Detection
of such a correlation allows one to study the role of
X-ray heating in the process of reionization. However, the
cross angular power spectrum of the CMB polarization and SXRB
maps constructed from ROSAT RASS and WMAP five-year maps is consistent
with no correlation. We attribute this negative detection to
both the extremely weak signals and large instrumental noises.
While a future search is needed with high sensitivity instruments
for both CMB polarization and soft X-ray photons, our current results can
still be used as a useful constraint on the effect of X-ray
heating in the epoch of reionization.
\end{abstract}


\section{Introduction}

Many efforts have been made so far to search for the correlation
between cosmic microwave background (CMB) and X-ray sky,
in an attempt to detect the signatures of both the integrated Sachs-Wolfe
effect resulting from time-evolving potentials of large-scale
structures at late times, which can be traced by various X-ray sources, and
the Sunyaev-Zel'dovich (SZ) effect due to the scattering of CMB photons
by the ionized electrons in clusters of galaxies.
Yet, no significant detections of the correlation signals
have been reported using the ROSAT RASS data \citep{kne97,die03},
in contrast to the positive results revealed by
tracers of large-scale structures, including
the hard (2-10~keV) X-ray background
\citep{bou93,ben93,ban96} and galaxies observed in the
 radio, optical and infrared bands
\citep[e.g.][]{scr03,bou04a,bou04b,nol04, mye04,fos04,afs04,pie06}.
In addition to the CMB-X-ray source correlation generated at late
times,  CMB polarization map may also correlate with
the unresolved soft X-ray background (SXRB) presumably
originated from supernovae,
X-ray binaries and quasars at high-redshift beyond $z=6$. These sources are
believed to play a significant role in the heating of neutral gas through
photoionization at the epoch of reionization \citep{dij04,che04},
while the ionized electrons generate the CMB polarization signals,
depending on both electron scattering optical depth and
primordial quadrupole of CMB. \citet{dor07} note that such polarization signals on small scales
are on order of $0.01~\rm{\mu K}$ \citep[see also][for theoretical prediction]{liu01} .
But as of today a quantitative estimate of the correlation signals is still
unavailable in the literature, and any of such detections will challenge the
existing instrumental limits.
The goal of this study is to make the first attempt at searching
for such a correlation, using the current upper limits of
the resolved diffuse SXRB and the recent WMAP polarization map.

A considerably large fraction  \citep[$\sim 94\%$, see][]{mor03} of soft and hard X-ray backgrounds have been
resolved into discrete sources, and the maximum admitted ranges of
the unresolved flux at different energy bands have been summarized
in \citet{wu01}. If internal physical processes are properly
treated,  diffuse emission of hot gas in groups and clusters
should also account for  the residual SXRB after the removal off
the discrete sources \citep{xue03}. This indeed restricts the room
for the presence of significant X-ray emission from high-redshift
sources at the epoch of reionization \citep[but see][]{sal07}. So the fraction
of soft X-ray which is emitted from high-redshift sources at the epoch of reionization
could be not more than $5\%$.
Furthermore, even if
the correlation is detected, the signature may have been interlaced with
those caused by
the SZ effect, in the sense that the CMB polarization anisotropies
may also be generated by hot electrons in clusters and even groups
of galaxies \citep{shi06}, unless foreground clusters and groups
including distant ones can be masked in the search.  In particular,
the fluctuations of SXRB in the epoch of reionziation may appear
at smaller scales comparable to or smaller than
the sizes of clusters, because X-ray photons usually have relatively
long free paths. This adds to further difficulties of our search
for the signatures of X-ray reionization.
Keeping these disadvantages in mind, we perform an analysis of
the correlation between the WMAP five-year polarization map and
ROSAT RASS archive  at energy band 0.2-2 keV, in the hope that
current limits are helpful for placing
robust constraints on models of X-ray reionization and for guiding
future works on this topic.

\section {Data}

For WMAP five-year maps,  we choose to  work with the V and W bands
in order to reduce foreground contamination. Furthermore, we
restrict the analysis to the differently assembly polarization data denoted
by V1, V2, W1, W2, W3, and W4, which are called WP sky maps hereafter.
Full five-year sky maps are obtained by performing a weighted,
pixel-by-pixel, mean of the single-year maps.
Polarization signals are characterized by the Stokes parameter Q, U, and V.
Because Thomson scattering can not generate circular polarization, we have
$V=0$. WMAP observes the sky with two orthogonal linear
polarization modes per feed, which are sensitive to the Stokes parameter
I, Q, and U \citep{hin07}. These data were released in
HEALPix\footnote{See http://healpix.jpl.nasa.gov} format
with different resolutions. The map used in this work is of Res 9
( $N_{\rm{side}}=512$, which means the angular resolution is 6.87' ).

The ROSAT RASS \citep{sno97} covers
the soft X-ray energy band ranging from $0.1$ to $2$~keV. The X-ray sky
in lower energies is severely contaminated by local emission~\citep{sli01}
while the X-ray sky in high energies shows a significant
contribution from extragalactic AGNs.
The RASS intensity ($I_{\rm SXRB}$) and noise ($\sigma$) maps are constructed
by $I_{\rm SXRB}=(C-B)/t_{\rm{exp}}$  and $\sigma=\sqrt{C+B}/t_{\rm{exp}}$, respectively,
where $C$ is the photon counts, $t_{\rm{exp}}$ is the exposure time and $B$ denotes the contamination,
which comprises
sources, particle background, scattered solar X-rays, long-term
enhancements, and so on.

We use two types of SXRB maps constructed from different algorithms.
The first one (hereafter RSI) is presented by \citet{sno97}.
RSI covers approximately 98\% of the sky in $1/4$ keV, $3/4$ keV,
and $1.5$ keV bands, each of which is further divided into two
bands (R1+R2, R4+R5, R6+R7, respectively) with an angular resolution of
about 12'. These maps were constructed using
an equal-area polar projection and divided into six areas according to
the six different projection pivots, in which
the contamination sources, such as the particle background , scattered solar
X-rays, and long-term enhancement, are excluded.

To produce an SXRB map with the angular resolution compatible to that of
WMAP data,  we construct the SXRB map (hereafter RSII map)
in HEALPix format with Res 9 ($N_{\rm{{side}}}=512$) from RASS archive
with energy band ranging from 0.2 to 2 keV,
in which each pixel covers
an area of $6.87'\times6.87'$ in the sky. Except for the correction to
exposure time, we make no attempt to remove the bright point and
diffuse sources from the local universe in order to optimize
 the detection of cross-correlation
with the WMAP polarization maps, even if it has a local origin.

\section {Statistical Method and Systematic Effects}

Following \citet{kog07}, we expand the polarization fluctuations
into generalized spherical harmonics
\begin{equation}
P^{(*)}(\mathbf{n})=Q(\mathbf{n})\pm iU(\mathbf{n})=
\sum a_{\mp,\ell m\ \mp 2}Y_{\ell m}(\mathbf{n}).
\end{equation}
The polarization fluctuations are conventionally decomposed into E-
and B-modes. Introducing a weighting function $w^P(\mathbf{n})$, we can write
the E-mode as
\begin{eqnarray}
\wt{a}_{\ell m}^{E} &=&\frac{1}{2}\int d\mathbf{n}w^P(\mathbf{n})[Q(\mathbf{n})
           (_2Y_{\ell m}^{*}(\mathbf{n})+_{-2}Y_{\ell m}^{*}(\mathbf{n})) \nonumber \\
      &&-iU(\mathbf{n})(_2Y_{\ell m}^{*}(\mathbf{n})-_{-2}Y_{\ell m}^{*}(\mathbf{n}))],
\end{eqnarray}
and its angular power spectrum is given by
\begin{equation}
  C_{\ell}^E=\frac{1}{2\ell+1}\sum_{m=-\ell}^{\ell}{a}_{\ell m}^E{a}_{\ell m}^{*E}.
\end{equation}

The fluctuation of SXRB intensity $\Delta I$  can be decomposed
in spherical harmonics as well
\begin{equation}
  \Delta I(\mathbf{n})=\sum_{\ell=0}^{\infty}\sum_{m=-\ell}^la_{\ell m}Y_{\ell m}(\mathbf{n}),
\end{equation}
and the corresponding angular power spectrum is
\begin{equation}
    C_{\ell}^I=\frac{1}{2\ell+1}\sum_{m=-\ell}^{\ell}a_{\ell m}^Ia_{\ell m}^{*I}.
\end{equation}
We define the cross-correlation power spectrum
between CMB E-mode map and SXRB as
\begin{equation}
  C_{\ell}^{x}=\frac{1}{2\ell+1}\sum_{m=-\ell}^{\ell}{a}_{\ell m}^Ea_{\ell m}^{*I}.
\end{equation}

To construct the angular power spectrum from the observations,
masks are often employed to account for the incomplete
or inhomogeneous sky coverages, or excisions of contaminated
regions, which are all incorporated in a window function $W(\mathbf{n})$.
For the WMAP polarization map, we choose the window function
$W_p(\mathbf{n})$ from the WMAP prescription \citep{ben03},
and we apply the natural weighting assumption to the window
function for the ROSAT map,  namely
$W_I(\mathbf{n})=0$ in the excised regions and $W_I(\mathbf{n})=1$ elsewhere.
The cross-power spectrum between SXRB and WMAP polarization maps
${C}_l^x$ in the presence of incomplete sky coverage is modified by
\begin{equation}
\wt{C}_{\ell}^x=\frac{1}{2\ell+1}\sum_m\wt{I}_{\ell m}\wt{E}_{\ell m}^*
\end{equation}
Such a pseudo cross-power spectrum $\wt{C}_l^x$ can be related to
the unbiased whole sky power spectrum through their ensemble averages
in terms of the algorithm MASTER
\citep{hiv02,han03,cha05,bro05}.
\begin{equation}
\langle\wt{C}_{\ell}^x\rangle=\sum_{\ell'}M_{\ell\ell'}\langle C_{\ell'}^x\rangle,
\end{equation}
where $M_{ll'}$ describes the mode-mode coupling
\begin{equation}
M_{\ell\ld} = \frac{(2\ld+1)}{4\pi} \sum_{\ldd}{\mathcal W}_{\ldd}^{IP}
\left( \begin{array}{ccc} \ell & \ld & \ldd \\
0 & 0 & 0 \end{array} \right)
\left( \begin{array}{ccc} \ell & \ld & \ldd \\
2 & -2 & 0 \end{array} \right),
\end{equation}
and $\mathcal{W}^{IP}$ is the cross-power spectrum of the window functions
\begin{equation}
\mathcal{W}_{\ell}^{IP}=\sum_mw_{\ell m}^I(w_{\ell m}^P)^*,
\end{equation}
with $w_{\ell m}^I$ and $w_{\ell m}^P$ defined as the spherical harmonic coefficients.

The angular power spectrum $\wt{C}_{\ell}^x$ should be
further corrected for the smoothing effects of both instrumental beams
($B$), finite pixels ($p$) and noise ($N$)
\begin{equation}
\langle\wt{C}_{\ell}^x\rangle=\sum_{\ld}M_{\ell\ld}B_{\ld}^IB_{\ld}^P
       p_{\ld}^Ip_{\ld}^P\langle C_{\ld}\rangle +\langle\wt{N}\rangle
\end{equation}
where $B_{\ld}^I$ and $B_{\ld}^P$ are the Fourier transforms of
beam patterns of ROSAT and WMAP, respectively. The pixel
window function in Eq. (11) is defined by
\begin{equation}
  p_{\ell}=\left(\frac{1}{N_{\rm pix}}\sum_{i=0}^{N_{\rm pix}-1}|p_{\ell}(i)|^2\right)
^{1/2},
\end{equation}
and the $m$-averaged window function reads
\begin{equation}
p_{\ell}(i)=\left(\frac{4\pi}{2\ell+1}\sum_{m=-\ell}^{\ell}|p_{\ell m}(i)|^2\right)
^{1/2},
\end{equation}
in which $p_{\ell m}$ is the spherical harmonic transform of pixel function
$P_{\rm pix}(\mathbf{n})=1/\Omega_{\rm{pix}}$, and $\Omega_{\rm{pix}}$ is the solid angle
subtended by one pixel.

Finally, the angular power spectrum should be properly binned to
reduce the systematic effects described above, and the associated
errors should also be estimated.  With the algorithms MASTER and
its extension to polarization, we can convert $\langle\wt{C}_{\ell}^x\rangle$
 to an 'unbiased' whole sky angular power spectrum $\mathcal{\wh{C}}$
and evaluate  the errors of
 $\mathcal{\wh{C}}$. In order to calculate the
errors and covariance of the $\mathcal{\wh{C}}$, we produce a number
of mock observational maps by Monte Carlo simulations.  So the elements of
the covariance matrixes are given by
\begin{equation}
\label{eq:err1}
\rm{C}_{bb'}=\left\langle\left (  \mathcal{\wh{C}}_b -\langle\mathcal{\wh{C}}_b\rangle_{MC}\right )
\left (\mathcal{\wh{C}}_{b'}-\langle\mathcal{\wh{C}}_{b'}\rangle_{MC}\right )\right\rangle_{MC}.
\end{equation}
The error bars on $\mathcal{\wh{C}}_b$ are then given by the square root of the diagonal
elements of $\rm{C}$
\begin{equation}
\label{eq:err2}
\Delta\mathcal{\wh{C}}_b=\rm{C}_{bb}^{1/2}.
\end{equation}

\section{Results}

Before we proceed to the study of the cross-correlation between
the CMB polarization map and SXRB, we present an extensive
analysis of the auto-correlation angular power spectrum of the two backgrounds
revealed by ROSAT and WMAP, respectively. This allows us to
have a better understanding of the statistical properties of our data sets.

We firstly work with the RSI maps in all the seven channels. Pixelizing
the RSI sky under the angular resolution of 12' gives rise to
a total pixels of $N_{\rm{side}}=256$ in the HEALpix format. In order to
match the WMAP angular resolution, the data are somewhat oversampled
so that a value of  $N_{\rm{side}}=512$ is actually used, which has
no influence on the construction of power spectra at large angular scales.
We have tested a few of weighting functions to compensate for
the incomplete sky coverage, and found that the results suffer from
the contamination of foreground structures, such as the
Galaxy and bright point sources in the soft X-ray band. We eventually
use the R6 data within $ b > 40^o $ and $ l \in [70,250]$,
the so-called MSR mask, which is suggested as a better representative
of the diffuse cosmological SXRB and less affected by the foreground
structures \citep{sli01}.
The angular power spectrum is constructed using
the HEALPix package  and the result is shown in Figure 1(a).

Next, we apply the same technique to the RSII map with pixel resolution
$N_{\rm{side}}=512$. Note that this sample has a wider
energy coverage ranging from $0.2$ to $2$~keV.
Since we make no attempt to edit the raw data except correct for the exposure
time, the RSII map looks very noisy and a large fraction of the sky
is apparently dominated by the emission of the Galaxy.
We demonstrate the resulting angular power spectrum with mask MSR
in Figure 1(b).

Despite the fact that the two samples have different energy coverages ,
the overall angular power spectra revealed by  RSI and RSII maps are
similar in shape.  Two samples differ by several orders of magnitude in amplitude because
the RSI maps divide up the flux of SXRB among seven energy  bands
,and only the power spectrum of R6 band map
is plotted in the Figure 1(a).  The slight excess in the RSII power spectrum at large scales
below $\ell\approx100$  can be attributed to the emission
from the galactic plane of the Galaxy.
Furthermore, the two angular power spectra both
have a steeper slope than $\ell(\ell+1)C_{\ell}\propto \ell^{0.8}$,
indicating that the SXRBs are actually dominated by
local sources (i.e., the Galaxy)
rather than X-ray galaxies distributed at cosmological distances.
Recall that the clustering of galaxies contributes an angular
power spectrum of roughly $\ell(\ell+1)C_{\ell}\propto \ell^{0.8}$,
provided that the sources follow a power-law angular power
spectrum with an index of $\sim0.8$ ~\citep{sco99}.
Yet, such a local component makes no contribution to the
cross-correlation of SXRB and CMB polarization to be explored below.

Now we turn to the CMB polarization sky manifested by the Stokes
Q and U parameters in the WP map. In a similar way to the analysis of
the SXRB maps above,
 we apply two masks in the analysis of CMB polarization map;
one is the full-sky WP data without any masks, and another is the
so-called P06 mask used by the WMAP team~\citep{pag07}.
For the latter, $25.7\%$ of the sky is masked near the Galactic
plane and around other strong sources of contamination, eliminating
the effect of synchrotron polarization by the Galaxy, dust,
point sources, etc.  Indeed, the angular power spectra of
polarization map with and without the P06 mask look rather different from
each other, as illustrated in Figure 2 using the CMB TE cross-power spectrum.
Apparently the structured power spectrum of the full-sky map originates
from the foreground, and the structures vanish immediately when the
contaminated regions of the sky are masked with the P06 mask. Shown also in
the Figure 2 is the theoretically predicted CMB TE power spectrum
calculated in terms of publicly available code
CMBFAST\footnote{http://cfa-www.harvard.edu/~mzaldarr/CMBFAST/cmbfast.html}.
It appears that the observationally measured and theoretically predicted
power spectra agree each other nicely when the P06 mask is applied.

Finally we carry out the study of cross-correlation between the CMB
polarization sky and SXRB using the ROSAT RSI and RSII samples and WP maps.
Construction of cross-power spectra can be made straightforwardly, and
the error bars of the cross-power spectra are estimated  from 100 mock WMAP observational
 polarization maps and 100 mock SXRB maps using Eq.~(\ref{eq:err1}) and Eq.~(\ref{eq:err2}).
The results of the cross-power spectra are shown in Figures 3 and 4 for RSI-WP and RSII-WP
cross-correlations, respectively, for which we have considered the sky maps
both
with and without masks in order to demonstrate to what extent the angular
power spectra are affected by the foreground sources.
For the WP maps all the six polarization databases
are employed and the results were averaged over the whole databases and the different
observational band are illustrated, respectively.
It turns out that, except for the RSII-WP cross-correlation without any masks,
the rest are essentially consistent with the negative
detection of cross-correlation between SXRB and CMB polarization maps.
Recall that even in the full-sky coverage without the application of
masks, such as MSR in the RSI (R6) sample, bright foreground X-ray sources have
already been excised. Employment of the MSR allows one to further
excise the Galactic disk from the samples. The absence
of significant correlations between the RSI (with and without masks)
or RSII (with mask MSR)  samples and WP maps may therefore be of
cosmological significance. We do detect a positive cross-correlation
in the RSII-WP full-sky maps without any masks over angular scales
of $\ell\approx 200-400$. Nevertheless, a careful examination of the RSII
full-sky map shows that such a positive correlation is related to three
strong X-ray emitting regions that are identified as Cyg Region,
 Gum Neb and Tau A
in the WMAP point source catalog, which are strong X-ray source and
supernova remnants.
 No correlation feature is detected
once these three regions are excised or MSR and P06 masks are utilized
(Figure 5). Actually, the result remains nearly unchanged if only one mask,
either MSR or P06, is adopted to all the cases.
The positive correlation associated with the three supernova
remnants can be interpreted as the polarization of synchrotron in the regions
or the polarization of CMB caused by the energetic
electrons there, yielding the cross-correlation between X-ray emission and CMB polarization map.

\section {Discussion and conclusions}
\label{dis}
The CMB polarization signals in large scale seen in the WMAP five-year data arise
primarily from the scattering of free electrons in the epoch
of reionization. At other scales, the primal quadrupole anisotropies
are still dominant sources.  But at small scales, the picture is more complicated. All types
of the perturbations beyond linear order will contribute to the polarization signals.
One possible source of polarization anisotropy at small scales is Thomson
scattering by free electrons at the epoch of the reionization. \citet{dor07} find
such polarization signals may have rms of order $\sim 0.01~\rm{\mu K}$ \citep[see also][]{liu01}.
Other secondary contribution to the signals includes
free electrons associated with large-scale structures
through the warm-hot intergalactic medium (i.e. missing baryons) in
local universe \citep{dav01}, and very energetic electrons
in clusters of galaxies \citep{bau03} . These secondary effects
($\sim 10^{-4}-10^{-5}\mu K$) are weaker than the polarization
signals produced in the epoch of reionization and believed to be
subdominant. Therefore, we focus on the
possible reason for our negative detection of cross-correlation between
ROSAT SXBR and WMAP CMB data.   Indeed, there is good reason to believe
that X-ray heating once playing a non-negligible contribution to
the reionization of the universe. Primary sources of X-ray emission
at high-redshift are high-mass X-ray binaries, supernovae and even
QSOs. It is natural to expect a positive correlation between
the CMB polarization and unresolved diffuse SXRB because both of them
originate from the same epoch and follow the common underlying
large-scale structures of the universe.

The most likely explanation
for the negative detection of the  cross-correlation
between CMB polarization map and SXRB obtained with
the WMAP and ROSAT maps may be due to the extremely faint signals
that are buried in the strong foreground and instrumentation noise.
Indeed, both CMB polarization and X-ray emission from the epoch of
reionization are rather weak,
and the latter has not even been separated from current SXRB. But if
we assume that the CMB polarization signals due to the inhomogeneous
process of the ionization of the universe
\citep[see][for the estimation of such CMB polarization signals]{dor07,liu01} were 100\% correlated with
the 5\% of the soft X-ray emission which is emitted from high-redshift sources at the
epoch of reionization, we can calculate this theoretical maximum correlations signals
and take it as a rough estimate for the actual correlations signals. The results of the estimate,
compared with the correlation results from the WMAP and ROSAT maps, are plotted as solid lines
 close to zero axes in Figure 3, 4, and 5.  It is quit evident that even the theoretical maximum
correlation signals are too weak to see in this measurement.
Furthermore, if most of the CMB polarization is generated in the later
time when the Ly$\alpha$ heating, instead of the X-ray heating, dominates
the process of reionization, the strength of the cross-correlation
between CMB polarization and SXRB would be even weaker. Therefore,
a quantitative estimate of the cross-correlation and its measurement
errors with existing and future instruments will be desired to explain
our negative detection ultimately, and to forecast further prospects.
The prime difficulty in carrying out such a theoretical investigation is
the uncertainty of the effect of X-ray heating sources and their
cosmic evolution in the epoch of reionization.

Obviously, such weak correlation
signals will provide an ambitious target for upcoming experiments.
Before a quantitative estimate is available, we can roughly evaluate the
detectability of the correlation signals with the theoretical maximum signals estimate.
Our rough estimation of the noise cross-power spectra for Planck
\footnote{http://www.rssd.esa.int/Planck}
($\sigma=2.3~\rm{\mu K}/5.0'$~fwhm, 0.85\% of the sky)
and ACT \citep[Atacama Cosmology Telescope;][]{kos06} or SPTPol
\citep[South Pole Telescope;][]{ruh04} ($\sigma=2.0~\rm{\mu K}/1.7'$~fwhm, 0.25\% of the sky)
polarization data correlating with ROSAT RASS data are illustrated in Figure 6, respectively.
The levels of the noise cross-power for CMB experiments like ACT or SPTPol  are above the theoretical maximum
signals levels, while those for Planck are under the theoretical maximum signal levels.
It appears that such theoretical maximum signals might be detected with Planck at $ \ell < 900 $.

In summary, if X-ray heating is one of the major sources to ionize the
neutral hydrogen in the epoch of reionization, we may expect to find a
positive correlation between the unresolved diffuse SXRB and CMB
polarization maps. Intuitively, our negative detection
based on the ROSAT and WMAP data indicates that such a correlation
is rather weak, and experiments with higher sensitivities are required to
make further explorations. Yet, our negative result can be used to
constrain the theory of X-ray heating in the epoch of reionization.
From our result, the theoretical prediction of the
cross power spectra between R6 band SXRB and CMB polarization
could not be out of range $-0.8 \lesssim \ell(\ell+1)/2\pi~C_{\ell} \lesssim 0.7~
\rm{(mK~10^{-6}cts/s/arcmin^2)} $
  at $\ell\simeq 300$,  $-1.5 \lesssim \ell(\ell+1)/2\pi~C_{\ell} \lesssim 1.2 $ at $\ell\simeq 500$,
 $-3.3 \lesssim \ell(\ell+1)/2\pi~C_{\ell} \lesssim 3.2 $ at
  $\ell\simeq 800$ (based on the result from RSI R6 band map and WP maps).
And those between wider energy band ($0.2~\rm{KeV}
\sim 2~\rm{KeV}$) SXRB and CMB polarization could not be out of range
  $-3.3 \lesssim \ell(\ell+1)/2\pi~C_{\ell} \lesssim 3.2~
\rm{(mK~10^{-6}cts/s/arcmin^2)} $ at $\ell\simeq 300$,  $-9.7 \lesssim \ell(\ell+1)/2\pi~C_{\ell} \lesssim 7.0$ at $\ell\simeq 500$,  $-36.4 \lesssim \ell(\ell+1)/2\pi~C_{\ell} \lesssim 30.2 $ at $\ell\simeq 800$
 (based on the result from RSII map and WP maps).
Meanwhile, a detailed study of  the cross
power spectrum would be is also worthy to guide
future observations.

We thank theWMAPand ROSAT teams for making their data
available via the Legacy Archive for Microwave Data Analysis
(LAMBDA) at http://lambda.gsfc.nasa.gov and the ROSAT XRay
All-Sky Survey archive at http://www.xray.mpe.mpg.de/
cgi-bin/rosat/rosat-survey. This work was supported by the
National Science Foundation of China (Grant no. 10673008)
and the NCET Program of Ministry of Education, China.

\begin{figure}
\plottwo{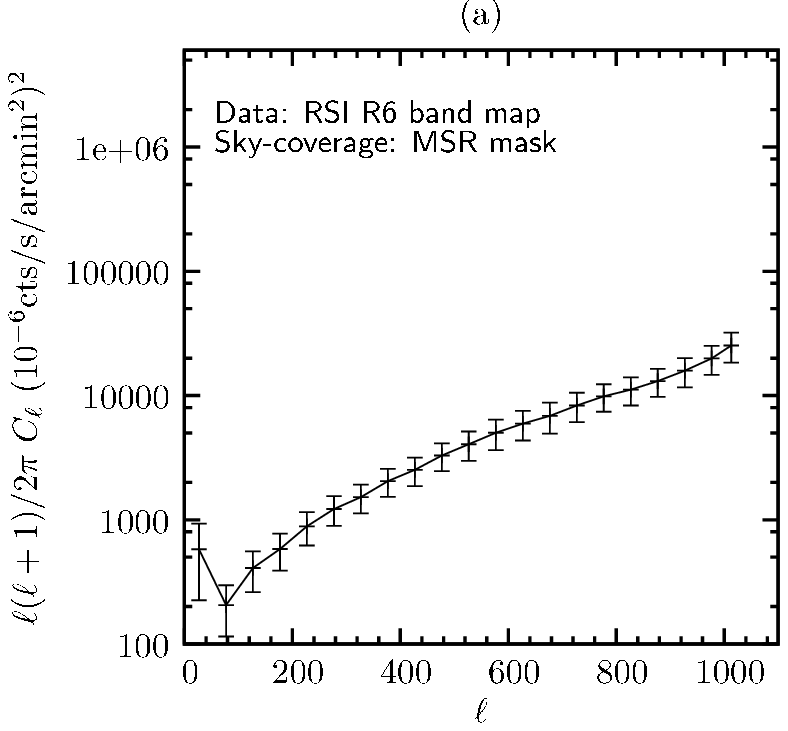}{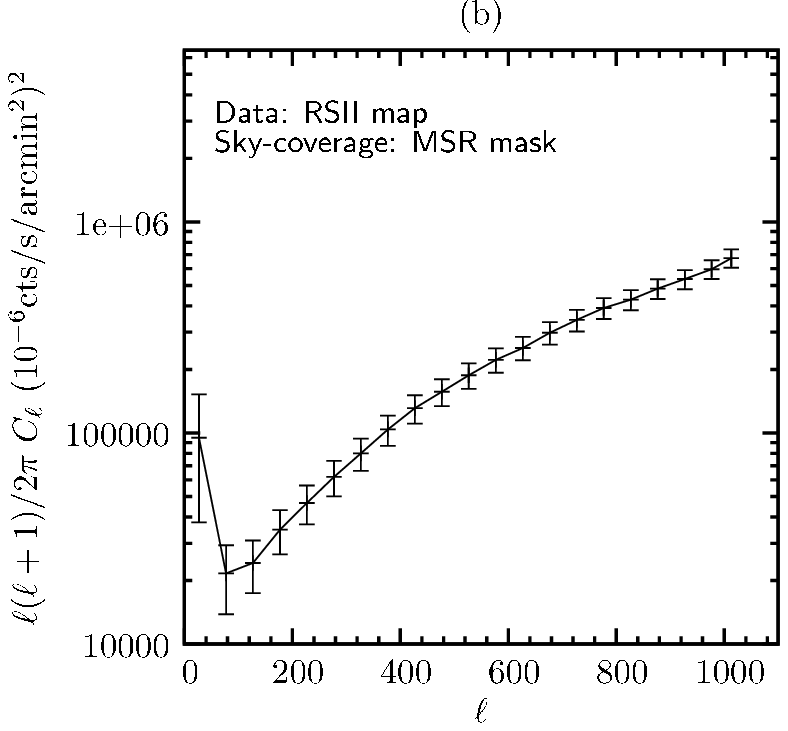}
 \caption{Angular power spectra revealed by the RSI and RSII samples.
The $1\sigma$ error bars are estimated using the WMAP simulations.
Both angular power spectra are binned in terms of $\Delta\ell=30$}.
\end{figure}

\begin{figure}
\plottwo{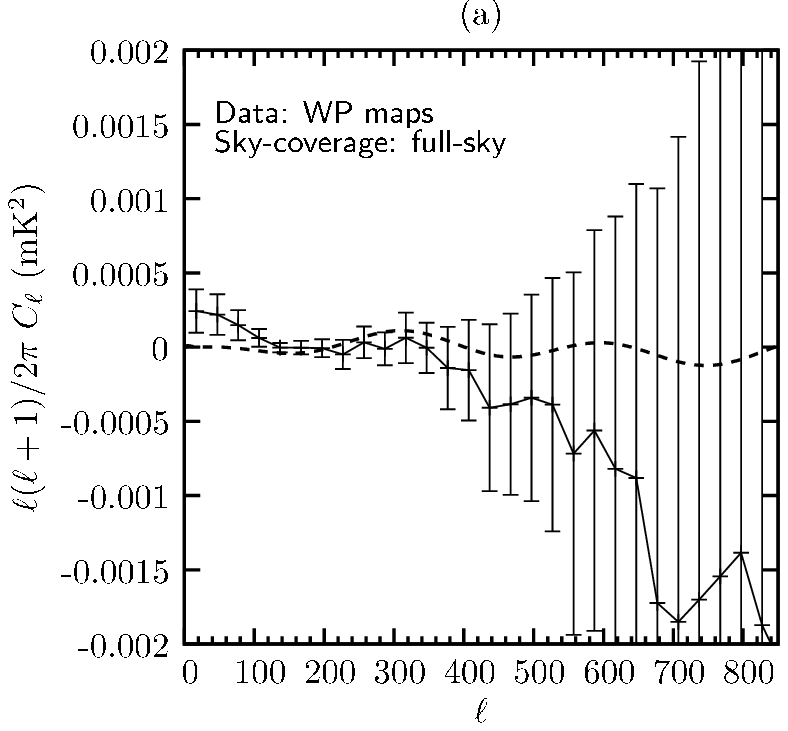}{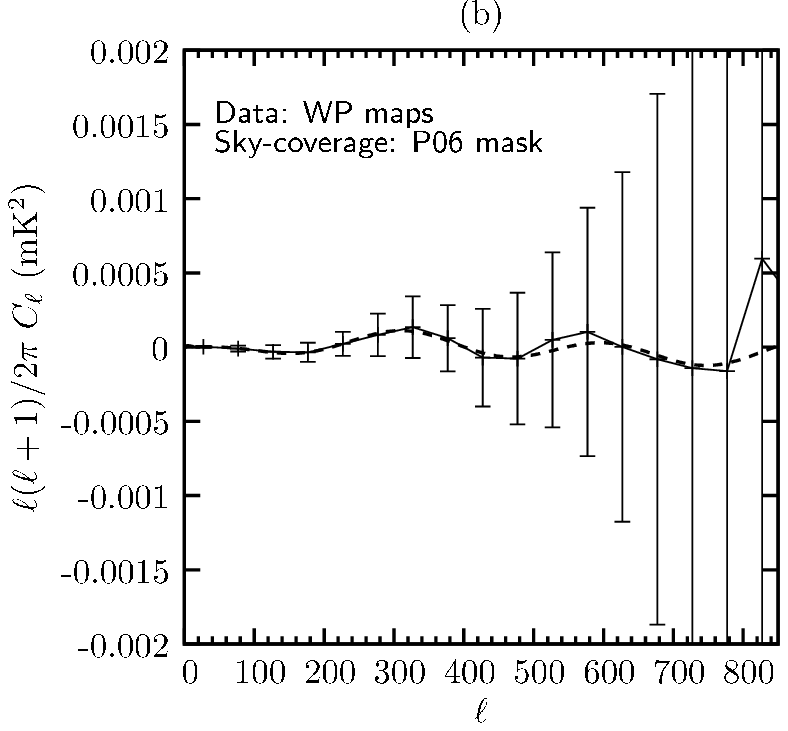}
\caption{TE angular power spectra constructed from WP maps(V1, V2, W1, W2, W3, W4
data)
with (b) and without (a) P06 mask, respectively.
The solid lines represent the results
obtained with the WP maps and the dashed lines are the theoretically
predicted power spectrum generated by CMBFAST. The angular power spectrum
constructed from the maps with masks are binned in terms of $\Delta\ell=50$.}
\end{figure}

\begin{figure}
\plottwo{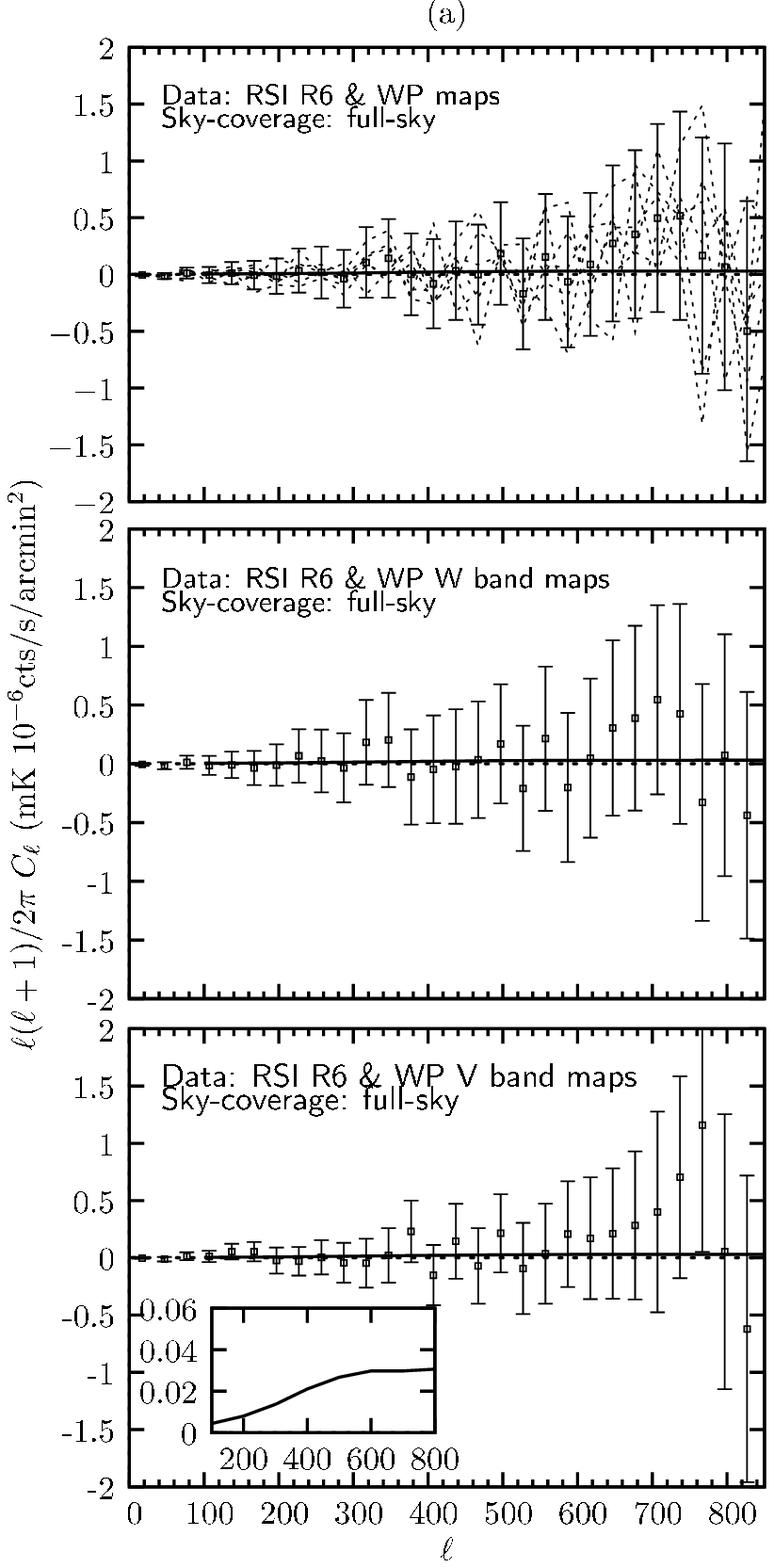}{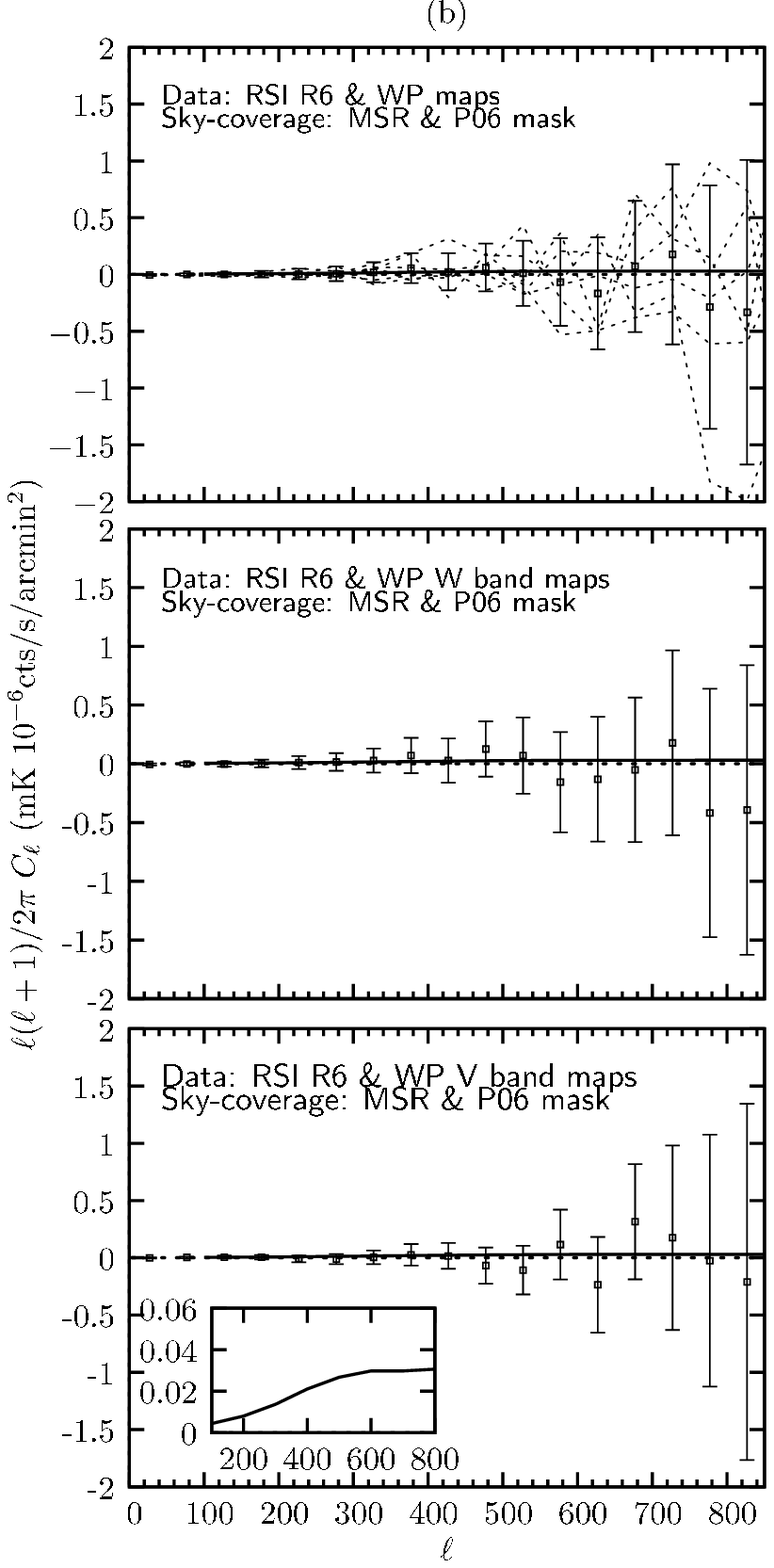}
\caption{Cross-power spectra constructed from RSI R6 sample and
WP six databases (dotted lines), and the open squares with error bars
represent their averages over the whole database ( top panel ), V ( middle panel ),
and W ( bottom panel ) band.  The solid line close to the zero axis represents the estimate of maximum
 correlation signals (see Section \ref{dis} for details).  The small graph sketches the estimate correlations
 signals with magnifying Y-range.}
\end{figure}

\begin{figure}
\plottwo{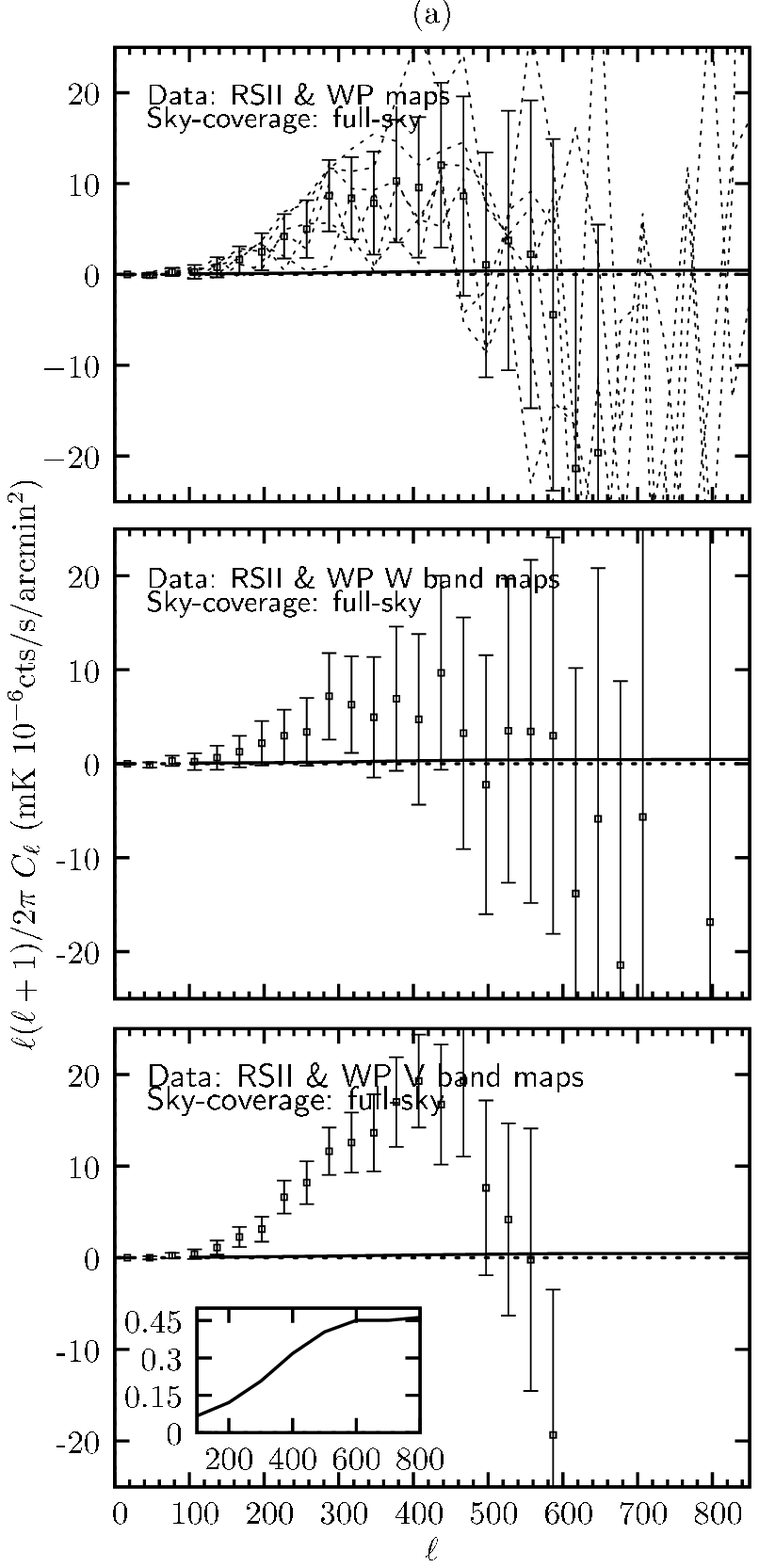}{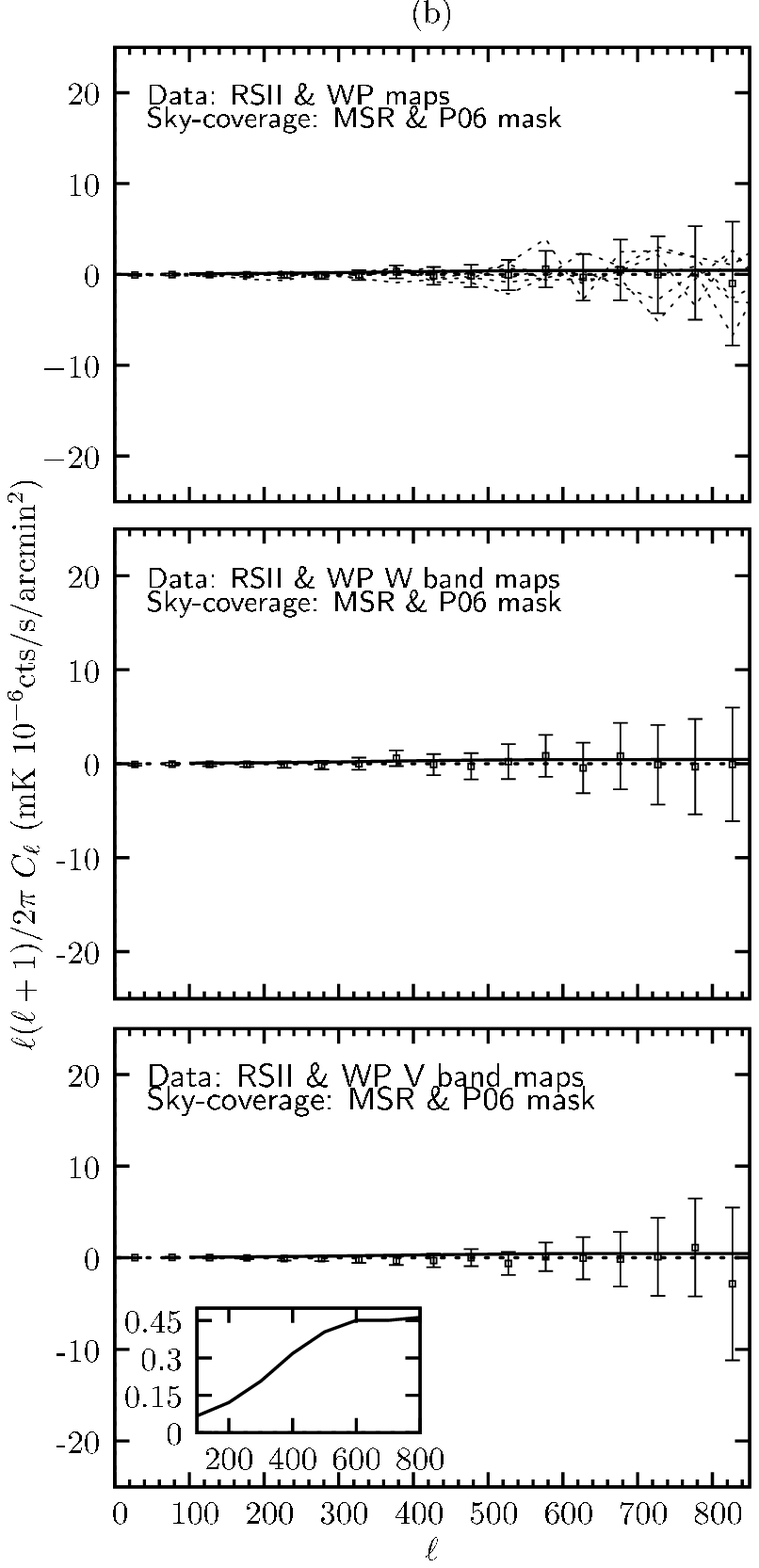}
\caption{Same as Figure 3, but for RSII sample. }
\end{figure}

\begin{figure}
\plottwo{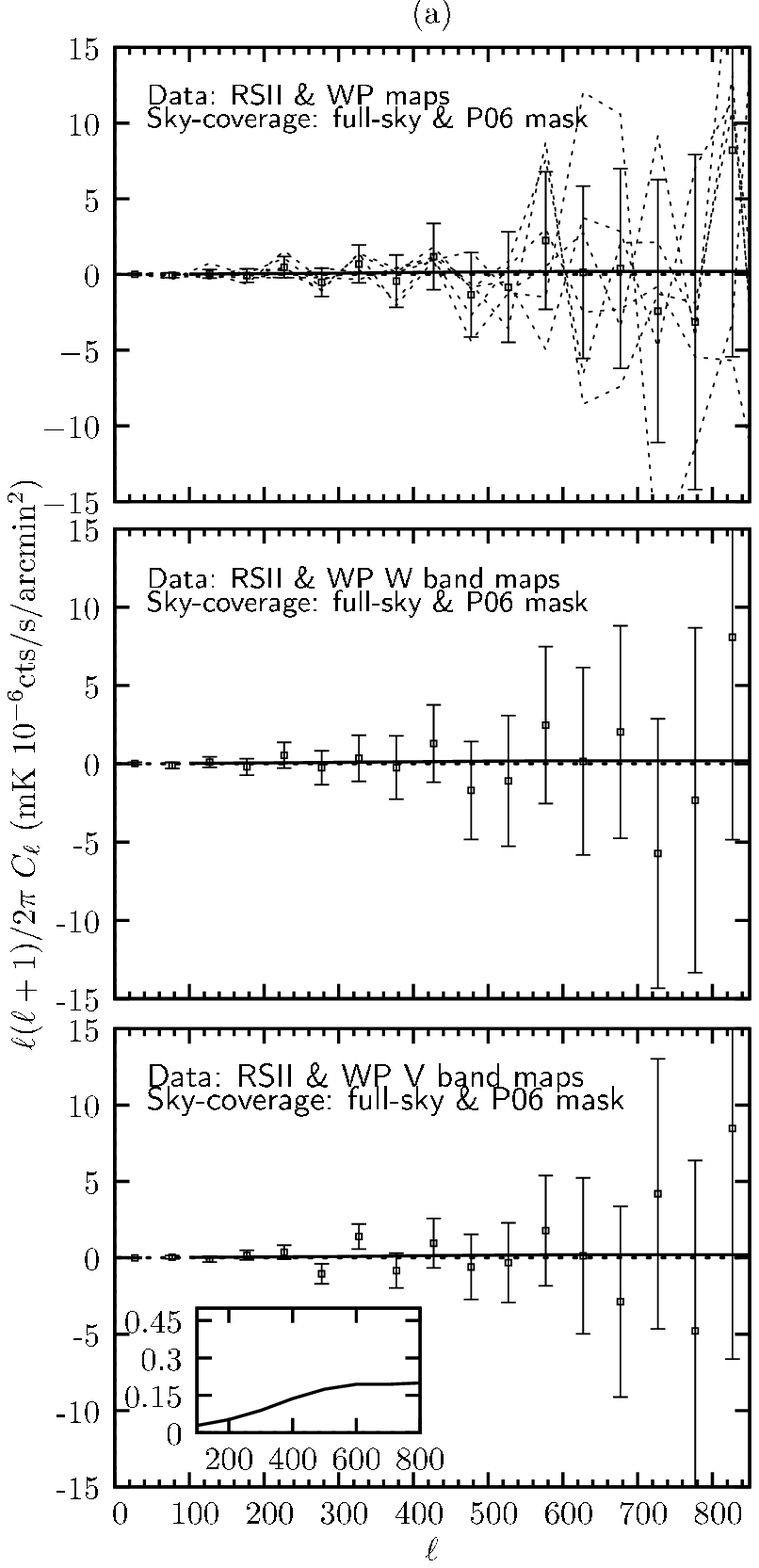}{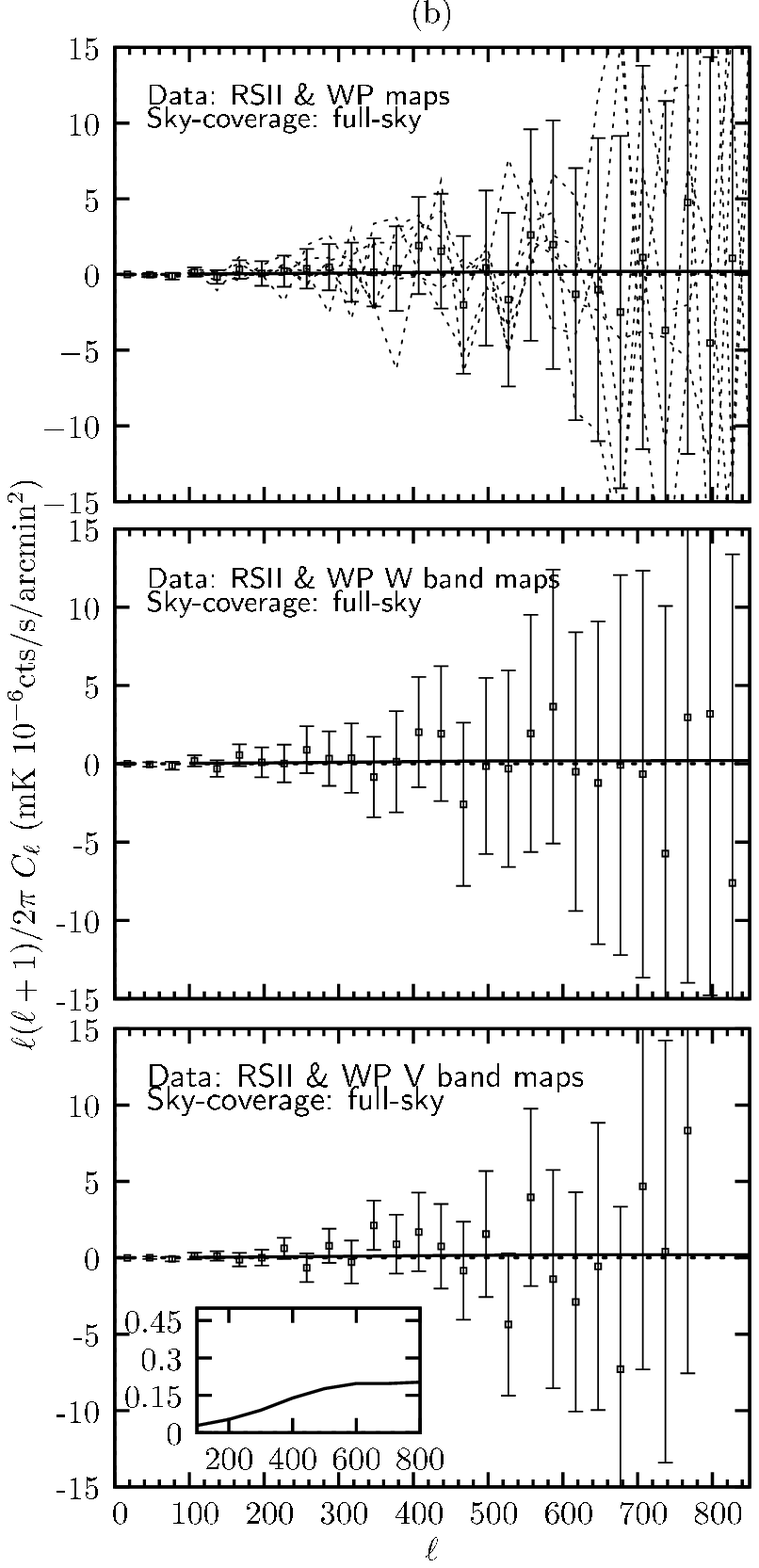}
\caption{Same as Figure 3, but for the full-sky RSII sample with
(a) the P06 mask on WP maps and (b) removal of three bright
X-ray sources in Galactic plane (Cyg Region, Gum Neb and Tau A).}
\end{figure}

 \begin{figure}
 \plottwo{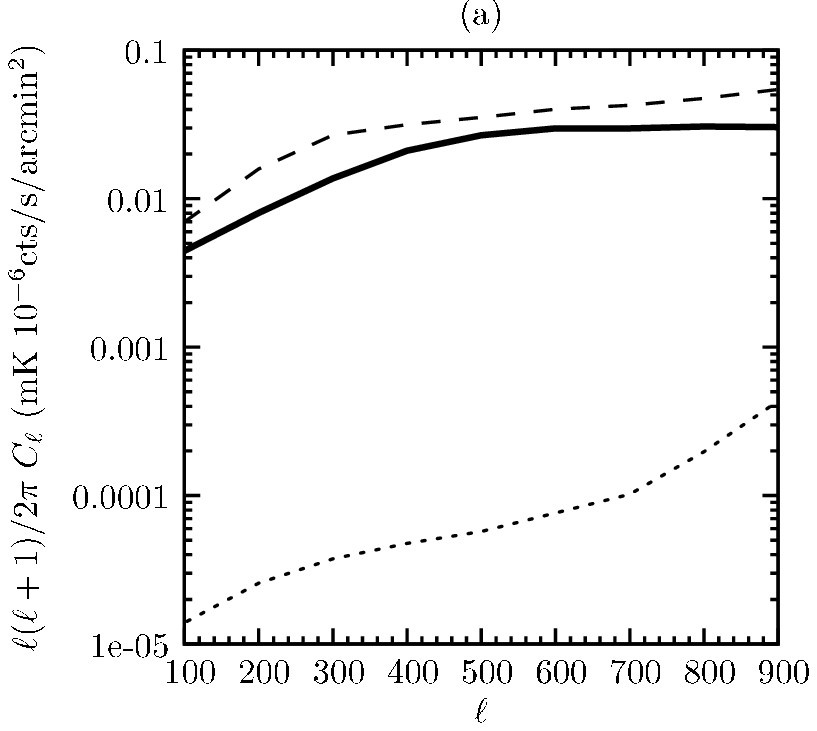}{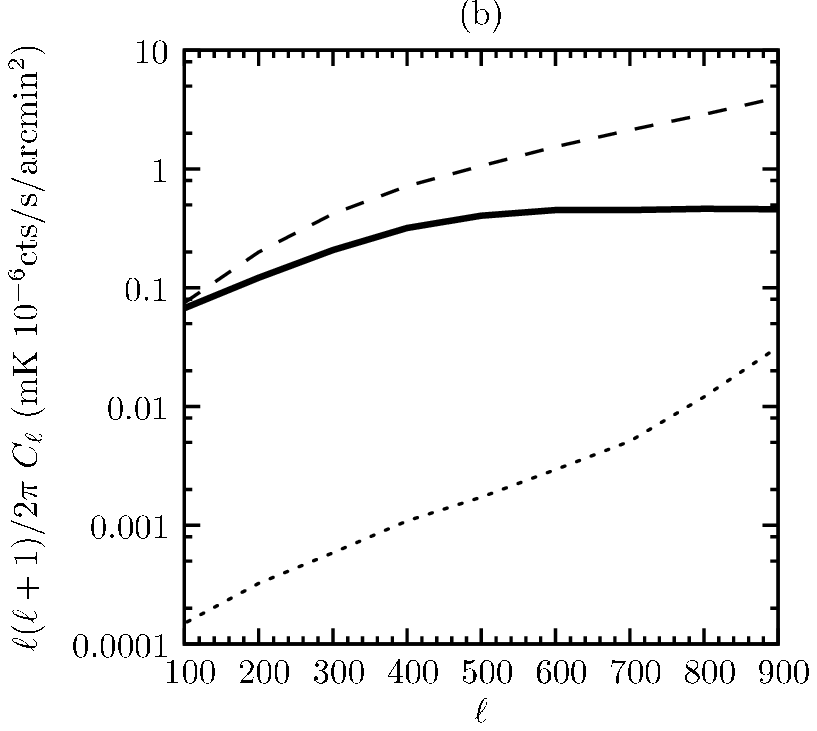}
 \caption{Theoretical maximum correlation signals one could expect compares with the estimated noise of
  the cross-power spectrum for the upcoming experiments. The heavy solid lines represent the estimated maximum
 signals from CMB polarization map correlating with RSI R6 band map (a)  and RSII map (b),  respectively. The
 estimation of the noise cross-power spectrum for Planck polarization and SXRB maps is shown as dotted lines,  while
 the estimation for the polarization observation like ACT or SPTPol  and SXRB maps is shown as dash lines.  }
\end{figure}

\end{document}